\documentclass[prl,amssymb,superscriptaddress, longbibliography,twocolumn]{revtex4-2}

\usepackage{amsmath}
\usepackage{amssymb}
\usepackage{mathtools}
\usepackage{mathrsfs}
\usepackage{xcolor}

\usepackage{graphicx}
\usepackage{dcolumn}
\usepackage{bm}
\usepackage{hyperref}
\hypersetup{colorlinks=true,citecolor={blue},linkcolor={blue},urlcolor={blue}}

\usepackage{color}

\begin{document}


\title{Quantum storage with flat bands}

\author{Carlo Danieli}
\affiliation{Institute for Complex Systems, National Research Council (ISC-CNR), Via dei Taurini 19, 00185 Rome, Italy}%

\author{Jie Liu}%
\affiliation{%
 College of Electronic Information and Physics, Central South University of Forestry and Technology, Changsha 410004, China
}%

\author{Rudolf A.\ R\"{o}mer}
\email{r.roemer@warwick.ac.uk}
\affiliation{
 Department of Physics, University of Warwick, Gibbet Hill Road, Coventry, CV4 7AL, UK
}%

\author{Rodrigo A.\ Vicencio}
\affiliation{%
Departamento de Física, Facultad de Ciencias Físicas y Matemáticas, Universidad de Chile, Santiago 8370448, Chile
}%
\affiliation{
Millennium Institute for Research in Optics - MIRO, Santiago, Chile
}

\date{\today}

\begin{abstract}
The realization of robust quantum storage devices relies on the ability to generate long-lived, spatially localized quantum states. In this work, we introduce a method for the targeted creation of compact excitations in flat-band lattices. By injecting \emph{in-plane} radiation waves from the system’s edge and applying a localized on-site potential at the desired storage position, we induce hybridization between compact localized states (CLSs) of the flat band and resonant dispersive plane waves. This hybridization enables the formation of spatially compact, stable excitations suitable for quantum memory applications. We experimentally validate this mechanism using photonic waveguide arrays, focusing on two representative geometries: the diamond chain and the one-dimensional Lieb ladder. Our approach is broadly applicable to any platform supporting flat-band physics.
\end{abstract}

\maketitle



Storage and retrieval of quantum states — also known collectively as {\it quantum memory} \cite{Simon2010QuantumMemories,Bussieres2013ProspectiveMemories,Saglamyurek2011BroadbandPhotons,Jiang2023QuantumCrystal,Lvovsky2009OpticalMemory} — are key building blocks of novel quantum technology architectures, such as quantum computers \cite{Lee2021ModelsReview,Gill2025QuantumChallenges,Memon2024QuantumBreakthroughs}, quantum communication \cite{Jiang2023QuantumCrystal,Radnaev2010AConversion,Wallucks2020AWavelengths,Ritter2012AnCavities,Hofmann2012HeraldedAtoms,Yu2020EntanglementKilometres,Usmani2012HeraldedCrystals,Moehring2007EntanglementDistance,Pu2021ExperimentalSegments}, and the quantum internet \cite{Wehner2018QuantumAhead,Kimble2008TheInternet}. Quantum memories use the quantum mechanical superposition principle to hold an exponentially growing number of states \cite{Bernien2013HeraldedMetres,Humphreys2018DeterministicNetwork,Chou2007FunctionalNetworks}. However, due to the non-cloning theorem \cite{Wootters1982ACloned,Dieks1982CommunicationDevices}, the readout of such quantum memories remains a fundamental conceptual challenge. 

On the other hand, the storage of classical information using single quantum states is much more advanced. Since such systems do not use entanglement to store information, there is no need to clone the full quantum state, and only the projected information \cite{Rontgen2019}, i.e., whether the state was occupied or not, is enough to read the classical bit information. Even such more simple quantum devices offer exciting potential for improvements over conventional storage devices \cite{Rontgen2019,Li2025RealizationLattice}. For example, modern sub-15 nm, capacitor-based DRAM devices use 10k - 60k electrons to store a single bit \cite{MicronDRAM}, while using a single electronic state should, in principle, suffice. Of course, such an approach then runs into similar decoherence and stability issues as for qubits -- unless one can find realisations of quantum states which are naturally stable and accessible.

In optical systems, such storage of light quanta (photons) has long been attempted. Kessel and Moiseev \cite{Kessel} discussed storage in a single photon state \cite{Mohan1998DelayedSelf-interference}. The experiment was demonstrated in 2003 \cite{Ohlsson2003DelayedDomain}. Optical data storage can be achieved by using absorbers \cite{LandyDavidR.2013AAbsorber,Turro2009PhotochemicalApplications}, e.g., different frequencies of light \cite{Munk2000FrequencyDesign}, which are then directed to beam space points and stored. Furthermore, light can be stored by conversion into an exciton and the life-time of the exciton can be enhanced by suitable engineering of the electron/hole separation \cite{Romer2000,Fischer2009a,Teodoro2010}, for example, in nano-ring geometries \cite{Fomin2018PhysicsRings}.

The principal requirement of such single-state quantum storage devices is then to identify a system of similar quantum states which are numerous -- to allow for storage of many classical bits -- while also spatially local to allow ready access for read-in and read-out. Furthermore, we want systems with good stability to maximise the lifetime of these states. Flat band (FB) lattices have many of these characteristics: they exhibit a macroscopically degenerate number of spatially compact eigenstates, that is, there are many of these states, and they are long-lived~\cite{Leykam2018ArtificialExperiments,Leykam2018Perspective:Flatbands, VicencioPoblete2021PhotonicDynamics,Danieli2024FlatApplications}. Experimentally, these locally compact flat band states have been generated in a diverse range of systems, from atomic gases \cite{Shen2010SingleLattices,Apaja2010FlatLattice,Taie2015CoherentLattice}, to solid-state devices \cite{Vidal1998Aharonov-BohmStructures,Abilio1999MagneticNetwork,Drost2017TopologicalLattices,Slot2017ExperimentalLattice}, and photonic lattices~\cite{Vicencio2015a,Mukherjee2015ObservationLattice,Xia2016DemonstrationLattices}. However, due to their degeneracy, the exact location of the compact eigenstates to be excited has thus far been determined largely by chance, when using non local means.

In this work, we present a mechanism to selectively excite compactly localised states (CLS) at desired positions, in two exemplary one dimensional flat band lattices: the diamond and the Lieb ladder. By shaping the local potential neighbourhood, we guarantee the excitation of CLS with the right phase structure. This aspect is particularly critical as, to the best of our knowledge, there is no any simple way of exciting a FB compact state at an arbitrary position by non-local means. We experimentally implement this approach in a photonic setup and show quantitative agreement with our theoretical quantum storage mechanism. The fabricated lattices prove to be readily controllable via in-plane injection of light from the boundary, for storage of classical bits at desired bulk plaquettes. The excitation method is based on a plane wave (PW) generator scheme,  which allows a selective excitation of specific radiation waves that resonate with the CLS at very precise regions.


\begin{figure}[t!]
	\centering
    \includegraphics[width=\columnwidth]{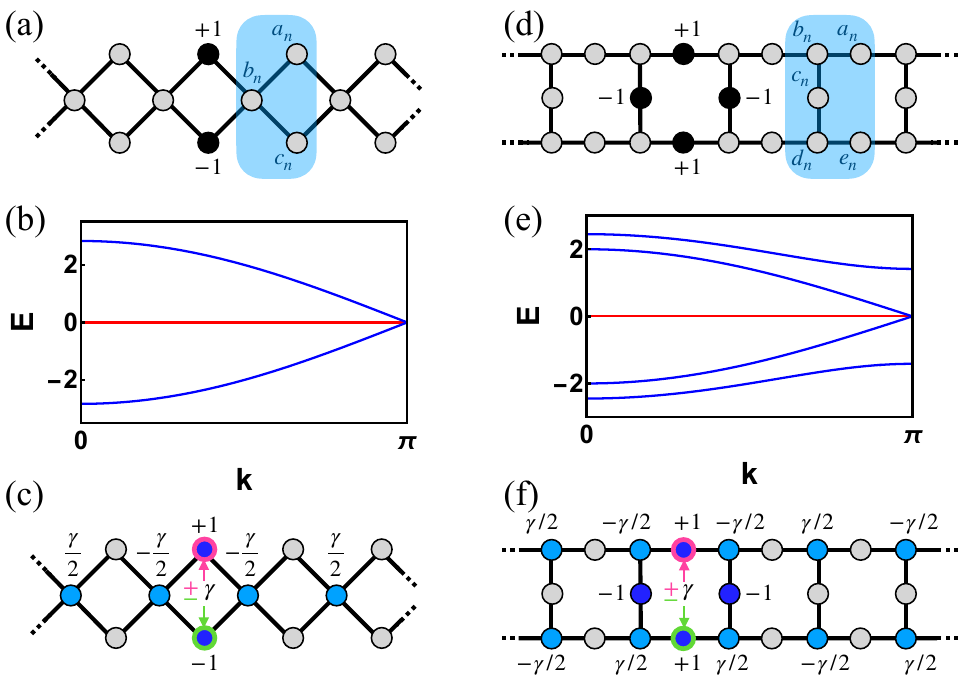}  
	\caption{
		(a) Schematic of the diamond chain. The three sites unit-cell is highlighted with the blue shaded area. The black circles show the position of the CLS with the correspondent amplitudes and phases as indicated. 
        (b) Band structure of the diamond chain. The flat band is colored in red, while the two dispersive bands are colored in blue.
        (c) Hybridized state at $E=0$. The blue circles indicate the location of the state with the correspondent amplitudes and phases as indicated. The onsite potentials are represented by the colored perimeters  $\gamma$ (magenta) and $-\gamma$ (green). 
        (d-f) Similar diagrams as (a-c) for the Lieb ladder.   
}
	\label{fig:1} 
\end{figure} 

Flat band systems are usually studied in the language of electronic tight-binding lattices and we shall follow this pathway here. For convenience, we study two paradigmatic models, namely, the so-called diamond chain and the quasi-1D Lieb ladder as shown schematically in Figs.~\ref{fig:1}(a) and (d). In both lattices, the sites are arranged in plaquettes of $M=4$ sites (for the diamond) and $M=8$ sites (for the Lieb ladder), with constant hopping strength (equal to one for simplicity). Both models exhibit a flat band at $E=0$, as detailed in Figs.~\ref{fig:1}(b) and (e), with corresponding CLS given in Figs.~\ref{fig:1}(a) and (d). They also possess, respectively, additional $\nu-1=2$ and $\nu-1=4$ dispersive bands for a total of $\nu=3$ and $5$ Bloch bands~\footnote{the dispersive bands of the diamond chain are  $E_{2,3}(k) = \pm 2\sqrt{2} \cos(k/2)$, while in the Lieb case the two dispersive bands $E_{2,3}$ of the diamond are paired up with  $E_{4,5}(k) = \pm \sqrt{2} \sqrt{2+ \cos k }$.}. We also note that these lattices are {\it chiral} -- namely, the lattices can be split in \emph{majority} $\mathcal{M}$ and \emph{minority} $m$ sublattices with a different number of components~\cite{Ramachandran2017ChiralStability,Calugaru2022GeneralBands}. For the diamond lattice, we have $\mathcal{M}_D = \{a_n,c_n\}_{n\in\mathbb{Z}}$ and $m_D=\{b_n\}_{n\in\mathbb{Z}}$, while for the Lieb ladder $\mathcal{M}_L = \{a_n,c_n,e_n\}_{n\in\mathbb{Z}}$ and  $m_D=\{b_n,d_n\}_{n\in\mathbb{Z}}$. The CLS associated with the flat bands at $E=0$ have profiles as shown in Figs.~\ref{fig:1}(a) and (d) by black dots, and they sit in the majority sublattices $\mathcal{M}_D$ and $\mathcal{M}_L$ in single plaquettes of the both systems. For the diamond chain the CLS form an orthogonal set of eigenstates of the flat band, while for the Lieb the CLS form a non-orthogonal basis \cite{Danieli2024FlatApplications}.

Flat band eigenstates are degenerate at $E=0$ for the lattices shown in Fig.~\ref{fig:1}. A travelling wave, resonant with the flat band energy, will move across the lattice without exciting a pre-selected CLS location. A way to generate such a selection comes from observing in Figs.~\ref{fig:1}(b) and (e) that the flat bands touch the dispersive bands at $k=\pm \pi$. This implies that there exist extended zero-energy states $\mathcal{E}$, which are confined in the minority sublattices $m_D$ and $m_L$ of the networks -- namely, $b_n=(-1)^n$ for the diamond chain and $b_n=(-1)^n$, $d_n = (-1)^{n+1}$ for the Lieb ladder. We then hybridize a number of target CLS at unit-cells $\mathcal{S} = \{n_j \}_{j=1}^\mathcal{N}$ with these $\mathcal{E}$ states by fine-tuning asymmetric onsite potentials at each $n_j$.

The tight-binding equations of motion of the lattices are 
\begin{equation}
i \dot{\Psi}_n 
=  \mathbf{H}_0\Psi_n +  \mathbf{H}_1\Psi_{n+1} +  \mathbf{H}_1^\dagger\Psi_{n-1} 
+ \gamma \sum_{n_j\in \mathcal{S} } \mathbf{V} \Psi_{n_j} \delta_{n,n_j} , 
\label{eq:fb_eq_P}
\end{equation} 
where $\Psi_n$ are complex vectors of $\nu$ components representing one unit-cell. The matrices $\mathbf{H}_0$, $\mathbf{H}_1$ and $\mathbf{V}$ are $\nu\times\nu$ matrices such that $\mathbf{H}_0$ defines the unit-cell profile, and $\mathbf{H}_1$ the hopping between neighboring unit-cells.
For the diamond chain, we have
\begin{equation}
\mathbf{H}_0 = 
{\small  \begin{pmatrix}
0 & 1 & 0 \\
1 & 0 & 1 \\
0 & 1 & 0 
\end{pmatrix}}, 
\quad
\mathbf{H}_1 = 
{\small \begin{pmatrix}
0 & 1 & 0 \\
0 & 0 & 0 \\
0 & 1 & 0 
\end{pmatrix}} ,
\label{eq:diamond_M}
\end{equation} 
while for the Lieb ladder
\begin{equation}
\mathbf{H}_0 = 
{\small \begin{pmatrix}
0 & 1 & 0 & 0 & 0 \\
1 & 0 & 1 & 0 & 0 \\
0 & 1 & 0 & 1 & 0 \\
0 & 0 & 1 & 0 & 1 \\    
0 & 0 & 0 & 1 & 0 \\    
\end{pmatrix}} ,
\ \ \text{and} \ \
 \mathbf{H}_1 = 
{\small \begin{pmatrix}
0 & 1 & 0 & 0 & 0 \\
0 & 0 & 0 & 0 & 0 \\
0 & 0 & 0 & 0 & 0 \\
0 & 0 & 0 & 0 & 0 \\
0 & 0 & 0 & 1 & 0
\end{pmatrix}} .
\label{eq:Lieb_M}
\end{equation} 
The matrix $\mathbf{V}$ defines the asymmetric potential localized within a unit-cell $n_j$, while the parameter $\gamma$ controls its potential strength. 
In our lattices, the matrices respectively are
    \begin{equation}
    \mathbf{V}_\text{D} = 
   {\small  
   \begin{pmatrix}
    1 & 0 & 0 \\
    0 & 0 & 0 \\
    0 & 0 & -1 
    \end{pmatrix}
    } ,
    \quad
    \mathbf{V}_\text{L} = 
     {\small    \begin{pmatrix}
    1 & 0 & 0 & 0 & 0 \\
    0 & 0 & 0 & 0 & 0 \\
    0 & 0 & 0 & 0 & 0 \\
    0 & 0 & 0 & 0 & 0 \\
    0 & 0 & 0 & 0 & -1
    \end{pmatrix}}
    \label{eq:Vmt_D_L} .
    \end{equation} 
The asymmetric potentials are visually represented in Figs.~\ref{fig:1}(c) and (f) by magenta rings (for positive $\gamma$) and green rings (for negative $\gamma$). These local potentials modify the CLS at unit-cell $n_j$, in both lattices, hybridizing them with the extended states $\mathcal{E}$.  This results in defect eigenstates at $E=0$ which we show in Figs.~\ref{fig:1}(c) and (f). In both panels, the dark blue sites in the perturbed plaquette have amplitudes equal to $\pm 1$, while the light blue sites have amplitudes $\pm \gamma/2$. For weak potential strength $\gamma\ll 1$, such impurity states resemble the unperturbed CLS on top of an oscillating extended background. These potentials can be introduced independently in several target unit-cells $\mathcal{S}$, yielding several defect states at $E=0$. 

The selection of one or more target CLS can now proceed via excitation of the hybridized states using an \emph{in-plane} incoming wave pulse. This novel mechanism in principle applies in all platforms described via a tight-binding representation, as in Eq.~\eqref{eq:fb_eq_P}, and supersedes the previous necessity of out-of-plane arrangements~\cite{Shen2010SingleLattices,Apaja2010FlatLattice,Taie2015CoherentLattice,Vidal1998Aharonov-BohmStructures,Abilio1999MagneticNetwork,Drost2017TopologicalLattices,Slot2017ExperimentalLattice,Vicencio2015a,Mukherjee2015ObservationLattice,Xia2016DemonstrationLattices}. We believe this to be an essential technical advancement for the engineering of future quantum storage devices. 
In the following, we explore this mechanism both theoretically and experimentally using optical waveguide lattices. In this case, the role of time $t$ is played by the propagation distance $z$ along the waveguides~\cite{Lederer2008DiscreteOptics}.

The quasi-bound-states-in-a-continuum (quasi-BIC's)~\cite{PedroBIC} generated by the asymmetric potential in Eq.~\eqref{eq:Vmt_D_L} are not orthogonal to zero-energy dispersive waves. Therefore they can be excited by launching an incoming pulse at $E=0$ from an edge of the lattice. 
This demands the excitation of a plane wave (PW) at a tailored quasi-momentum $k_0$ (in our cases, $k_0=\pi$) with a Gaussian-like spatial envelope which should be 
wide enough to have a narrow representation in momentum space around $k_0$.  This concept, although being a standard technique in theoretical physics~\cite{fanoBEC,fano2D}, is hard to be realized in practice~\cite{CANTILLANO2017}. However, very recently~\cite{Real2024Radiation-basedLattices} an experimental method to overcome this problem has been suggested, where the excitation of a side defect waveguide~\cite{SebaNL25} yields the existence of an impurity state with a flat extended oscillating tail. The sharpness of the PW generator is strongly dependent on the coupling between the side defect and the respective lattice, but also on the specific geometry~\cite{SM}.

\begin{figure}[t!]
	\centering
    \includegraphics[width=\columnwidth]{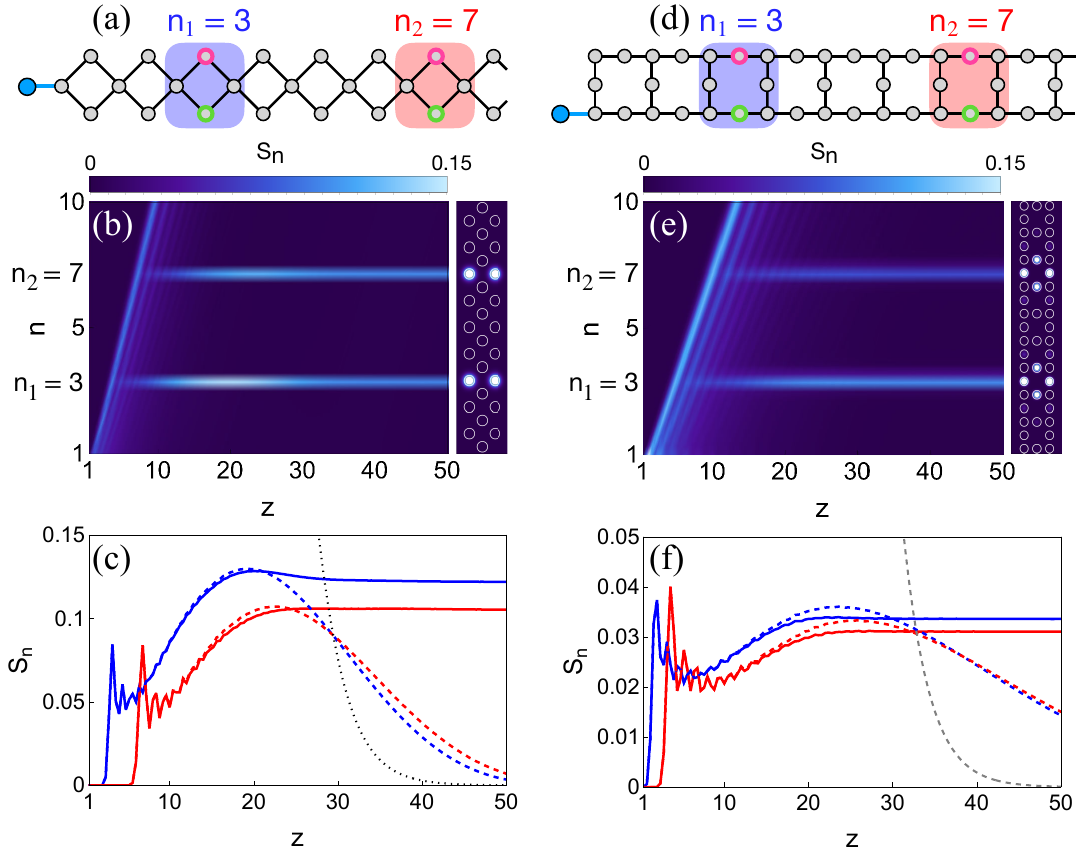}  
	\caption{
		(a) Eight plaquettes of the diamond chain. The two plaquettes with the onsite defects at $n=3$ and $n=7$ are highlighted with the blue and red shaded areas, respectively. 
		The dark blue circle is the defect waveguide used to inject the wave-packet around $E=0$. 
        (b) Intensity profile evolution $S_n$ for an initial single-site excitation centred at the defect waveguide.  The vertical right panel shows the local intensities $|\psi_n^j|^2$ at the final propagation distance.  
        (c) Intensity evolution $S_n$ for unit-cells $n_1=3$ (blue) and $n_2=7$ (red). The solid lines are obtained with the sigmoid potential release applied, while the dashed curves are without. The black dotted curve indicates the sigmoid curve.  
        (d-f) Similar diagrams to (a-c) for the Lieb ladder.   
}
	\label{fig:2} 
\end{figure}

In our lattices, we realize the PW method by attaching a defect waveguide $u$ in the left side of the lattices -- guide colored in blue in Figs.~\ref{fig:2}(a) and (d). This means that in the first unit-cell $n=1$ of the diamond chain, the central site is governed by  the equation $i\dot{b}_1 = a_1 + c_1 + V u$ for a coupling strength $V$, while the equation of the defect waveguide is $i\dot{u} =  V b_1$, as shown in Fig.~\ref{fig:2}(a). In the Lieb ladder instead, we implement the same protocol by attaching the defect waveguide $u$ to the bottom-left guide $d$ of the unit-cell, as shown in Fig.~\ref{fig:2}(d).

We numerically integrate the augmented lattice Eq.~\eqref{eq:fb_eq_P} for a single-site excitation in the defect waveguide $u(0) = 1$, $\Psi_n(0) = 0$, which introduces a radiating wave-packet in the middle of the dispersive bands $E=0$ for $k= \pi$~\footnote{Note that radiating wave-packet in different part of the dispersive bands can be achieved by introducing an onsite potential $\beta_u$ in the defect waveguide -- {\it i.e.} rendering its equation  $i\dot{u} = \beta_u u +  V_u b_1$.}. 
%
%
In our test, we perturb two unit-cells $\mathcal{S} = \{3,7\}$, as shown in Fig.~\ref{fig:2} in blue and red shaded areas, respectively. Note that here we present simulations using relatively small system sizes ({\it i.e.}, number of plaquettes) in order to compare with the experimental realization set-up. Clearly, our results extend to lattices of hundreds of sites and several perturbed unit-cells. 
In Fig.~\ref{fig:2}(b) we show the light intensity in each unit-cell $S_n(z)$ along the propagation distance $z$~\footnote{In the diamond chain $S_n(z) = |a_n(z)|^2 + |b_n(z)|^2 + |c_n(z)|^2$, while for the Lieb ladder $S_n(z) = |a_n(z)|^2 + |b_n(z)|^2 + |c_n(z)|^2 + |d_n(z)|^2 + |e_n(z)|^2$}. 
We turn off the impurities after the radiating pulse has passed by applying a sigmoid to the potential strength $\gamma$  -- {\it i.e.} the strength decays as  $\gamma(z) = \gamma_0 \big[ 1- \frac{1}{ 1+e^{-(z-z_0)/\tau} }  \big]$, with $\gamma_0 = 1/8$, $z_0 = 20$ and $\tau=2.5$. 
In Fig.~\ref{fig:2}(b), for $0\leq z\lesssim 10$ we observe an initial ballistic traveling pulse generated by the defect waveguide. For $z\gtrsim 10$, after the PW front propagated beyond unit-cell $n=10$, the only non-vanishing local intensities $S_n$ are at the perturbed unit-cells $n=3$ and $n=7$.  
The vertical strip next to the propagation plot shows the output intensity profile. We observe that the light intensities are trapped only in the $a$ and the $c$ sites of the perturbed unit-cells, analogously to the unperturbed CLS. 
Fig.~\ref{fig:2}(c) clarifies this by showing the intensities $S_n$ in the target unit-cells $n=3$ (blue) and $n=7$ (red) with solid lines, which saturate at finite values. 
These results indicate that the light intensities which are trapped in the $a$ and the $c$ sites of the perturbed unit-cells are not released since the connection with the lattice has been switched off via the sigmoid dependence. 
For comparison, in Fig.~\ref{fig:2}(c) we show with dashed lines that,  for a constant $\gamma = 1/8$,  the intensities $S_n$ in the target unit-cells $3$ and $7$ are decaying towards zero. Such a mechanism also occurs in the Lieb ladder, as shown in Figs.~\ref{fig:2}(d)-(f). In this case, differently from the diamond chain, the CLS form a non-orthogonal basis of the $E=0$ flat band and the impurity state also is non-orthogonal to its neighboring CLS. 
Hence, exciting the impurity state via the local defect in Eq.~\eqref{eq:Vmt_D_L} induces seemingly exponentially decaying excitations on neighboring plaquettes~\cite{Chalker2010AndersonBands}. 
As shown in Fig.~\ref{fig:2}(e), this also implies that the amplitudes in the $c$ sites, which induce the non-orthogonality, are less bright (weaker) than the $a$ and $e$ sites. Nevertheless, this non-orthogonality does not prevent the excitation of strongly localized states via an incoming plane wave.

    

\begin{figure}[t!]
	\centering
    \includegraphics[width=0.95\columnwidth]{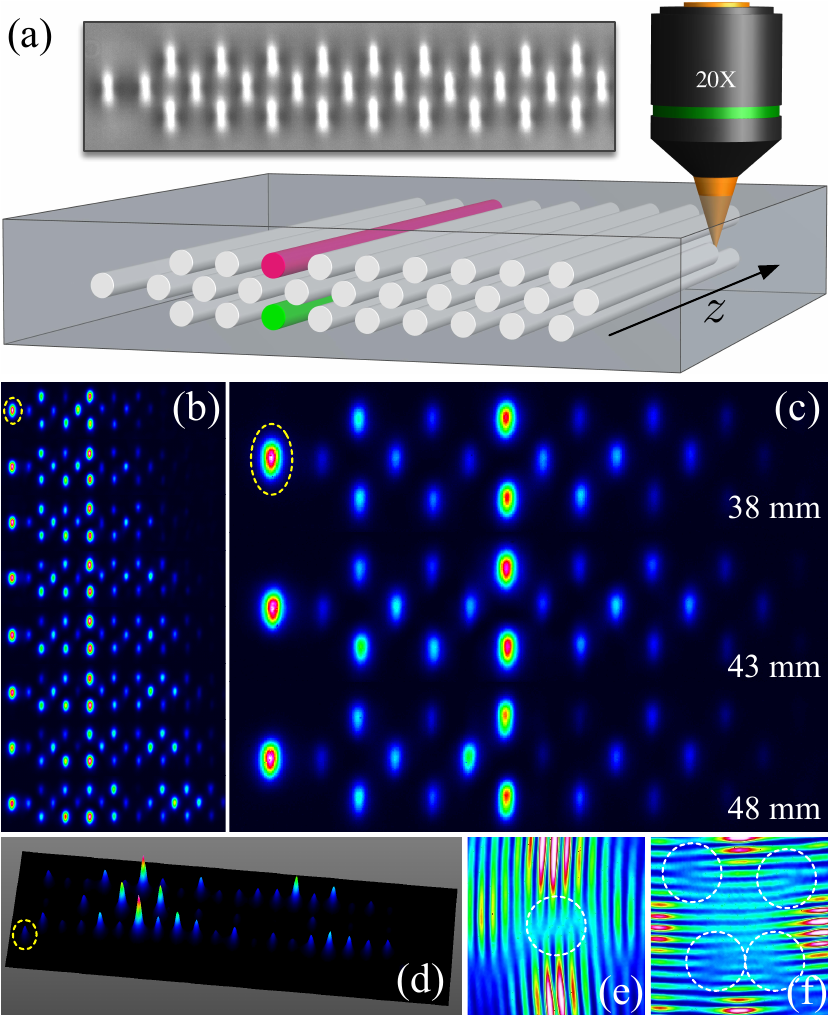}  
	\caption{
		(a) Sketch of the femtosecond laser writing technique describing the plane wave generator, a diamond lattice and the asymmetric potential. The (top) inset shows a microscope image of a fabricated diamond lattice. (b) and (c) Excitation of a FB localized state at the third ring of a diamond lattice for a $\lambda$-scan in the interval $\{720,790\}$ nm and for a $z$-scan of the asymmetric potential at $\lambda=745$ nm with $z$-distance as given in mm for (c). (d) Excitation of a localized FB state at the third ring of a Lieb ladder at $\lambda=765$ nm. (e) and (f) Interferogram at $\lambda=730$ nm for the excitation of diamond and Lieb ladders, respectively, with circles indicating the phase discontinuities. Yellow ellipses show the input position.}
	\label{fig:3} 
\end{figure}

We fabricate several 1D photonic lattices by means of the femtosecond (fs) laser writing technique~\cite{szameit_discrete_2005}, as sketched in Fig.~\ref{fig:3}(a). A fs laser is tightly focused inside a borosilicate glass chip ($n_0=1.48$) and weakly modifies the refractive index contrast in the order of $10^{-4}-10^{-3}$ over $n_0$. The magnitude of the final change is determined by the fs laser power and the writing velocity in which the automatized translation stage moves in the propagation coordinate $z$. In this specific experiment, we fix the nominal fabrication power to $P_w=20.5$ mW and the writing velocity to $v_w=1.0$ mm/s. The asymmetric potential is experimentally inserted by modifying the velocity of the upper and lower sites at a specific plaquette. A linear and well controlled dependence of the waveguides propagation constants is achieved in our setup around the chosen $v_w$~\cite{SM}, where we simply take a faster (slower) writing velocity for deeper (shallower) waveguides. The propagation constant in the photonic implementation is directly equivalent to the site energy of tight-binding models~\cite{repLederer,SPdimer}, and to the potential strength $\gamma$.

All our experiments were performed on a $L=50$ mm glass wafer, the maximum propagation length. However, we can effectively elongate this length by increasing the coupling constants through a wavelength-scan method~\cite{diamondAPL23,graphene24}. This allows us to adjust the effective propagation coordinate $z$, which in tight-binding like models can be written as ``$V z$'', with $V$ the coupling constant (hopping). As larger wavelengths $\lambda$ excite spatially wider guided modes, the coupling constants increase linearly with $\lambda$~\cite{2Drad25}. This means that a single lattice structure can be studied dynamically by ramping the wavelengths and by imaging the output intensity profiles. For example, Fig.~\ref{fig:3}(b) shows the effective evolution on a diamond lattice for a weak defect coupling, such that a significant part of the energy remains at the defect site. We observe how the excited wavepacket propagates to the right and excites the FB state at the third plaquette. This figure shows very clearly the efficiency of our scheme, observing a quite perfect localized state even for larger wavelengths. 
After finding the optimal experimental conditions for the PW generator on diamond chains and Lieb ladders~\cite{SM}, we look for optimizing the switching off of the asymmetric potential, such to trap the FB localized state on a fully FB lattice. Fig.~\ref{fig:3}(c) shows an optimization for the asymmetric potential, at the indicated lengths for the diamond chain. We observe an optimal result for $z=43$ mm, with a larger population of the FB state. These results confirm the numerical prediction using a sigmoid function for the inserted potential, so as to finally excite a pure FB lattice with a perfect FB out of phase profile.

We also performed several experiments on Lieb photonic ladders. The main challenge in this geometry is the increment of lattice sites and possible next-nearest neighbour couplings~\cite{desyat12,FBluis}, as well as the non-orthogonality of CLS in this geometry. In addition, the natural waveguide ellipticity of the fabrication method~\cite{szameit_discrete_2005} produces a strong asymmetry of vertical and horizontal coupling constants~\cite{SM}. This produces a very asymmetric FB state, with essentially two main peaks at the top and bottom sites of the respective plaquette. After an intensive optimization process~\cite{SM}, we were able to excite a quite clear symmetric Lieb mode at the third plaquette, as shown in Fig.~\ref{fig:3}(d). In this case, the defect site is well coupled to the lattice and almost no energy remains at the excitation position. 
Finally, Figs.~\ref{fig:3}(e) and (f) show the interferograms at $730$ nm for diamond and Lieb FB states, respectively. For diamond, we expect a FB state occupying the upper and bottom sites of a plaquette, with an out of phase configuration~\cite{DiamondSeba,diamondAPL23} [see Fig.~\ref{fig:1}(a)]. Fig.~\ref{fig:3}(e) shows an out of phase pattern with a clear fringe discontinuity in between these two sites. For Lieb ladders~\cite{1DLieb}, the FB state has four sites with a staggered phase structure~\cite{Vicencio2015a,DiamondSeba,FBluis} [see Fig.~\ref{fig:1}(d)]. This means that opposite sites are in phase, as we observe in Fig.~\ref{fig:3}(f) with a continuation of fringes. However, the staggered phase profile is expressed in the regions in between the four sites, observing in this figure clear fringe discontinuities, and a quite clear $X$-like phase profile. 


In conclusion, our numerical and experimental results show that the notion of using CLS as quantum storage devices is indeed possible. The ability of locally changing the environment to create an initial trap for the quantum state to slot into can be seen as a local quantum ``gate''. Once sufficient intensity has accumulated in the indicated plaquette, the trap can be released and the compactness of the FB states leads to a natural and perfect spatial confinement. The important conceptual advancement lies in the fact that the excitation of the state does not need to be done locally, but can rather be injected in-plane from the boundary of the device. This simplifies the engineering considerations considerably.
Of course, a readout mechanism is also required. This is readily available by the same trap mechanism. Simply turning on the local gate leads to a release of the state as the perfect local resonance conditions no longer hold. The intensity then leaves the CLS and can be measured as a pulse intensity when leaving the system.
While our demonstration has been done in photonic systems, the same principles hold for the full range of linear wave phenomena, from acoustics to electronics.
Furthermore, while our quasi-1D systems might seem restrictive, it should be clear that quasi-2D and -3D systems can be constructed by parallel lines and stacks of sheets thereof, respectively. The addition of external circuitry~\cite{Chang2021Symmetry-InducedLattice} could give extra degrees of freedom for a fully controllable and operational quantum storage device.

Last, one might want to speculate if the CLS mechanism can also support the concept of quantum memory based on entanglement as discussed above. When interactions -- or non-linearities -- are carefully controlled, flat-band systems such as the 3D Moire lattices might provide suitable hosts for such memories~\cite{Gao2025Low-dimensionalLattices}.

\begin{acknowledgments}
C.D., J.L.\ and R.A.R.\ thank the Center for Theoretical Physics of Complex Systems at the Institute for Basic Sciences, Daejoon, Korea for generous hospitality in the initial stages of this work. R.A.V.\ acknowledges financial support from Millennium Science Initiative Program ICN17$\_$012 and FONDECYT Grant No. 1231313. J.L.\ gratefully acknowledges funding support from the General Project of Hunan Provincial Education Department, China (Grant No.\ 23C0102) and the Hunan Provincial Natural Science Foundation (Grant No.\ 2024JJ6708).
UK research data statement: Data is available at https://wrap.warwick.ac.uk/192293/.
\end{acknowledgments}



%


\clearpage

\onecolumngrid

\section*{Supplementary material}

\subsection{Excitation's tails} 

\begin{figure}[h!]
  \center{\includegraphics[width=\columnwidth]{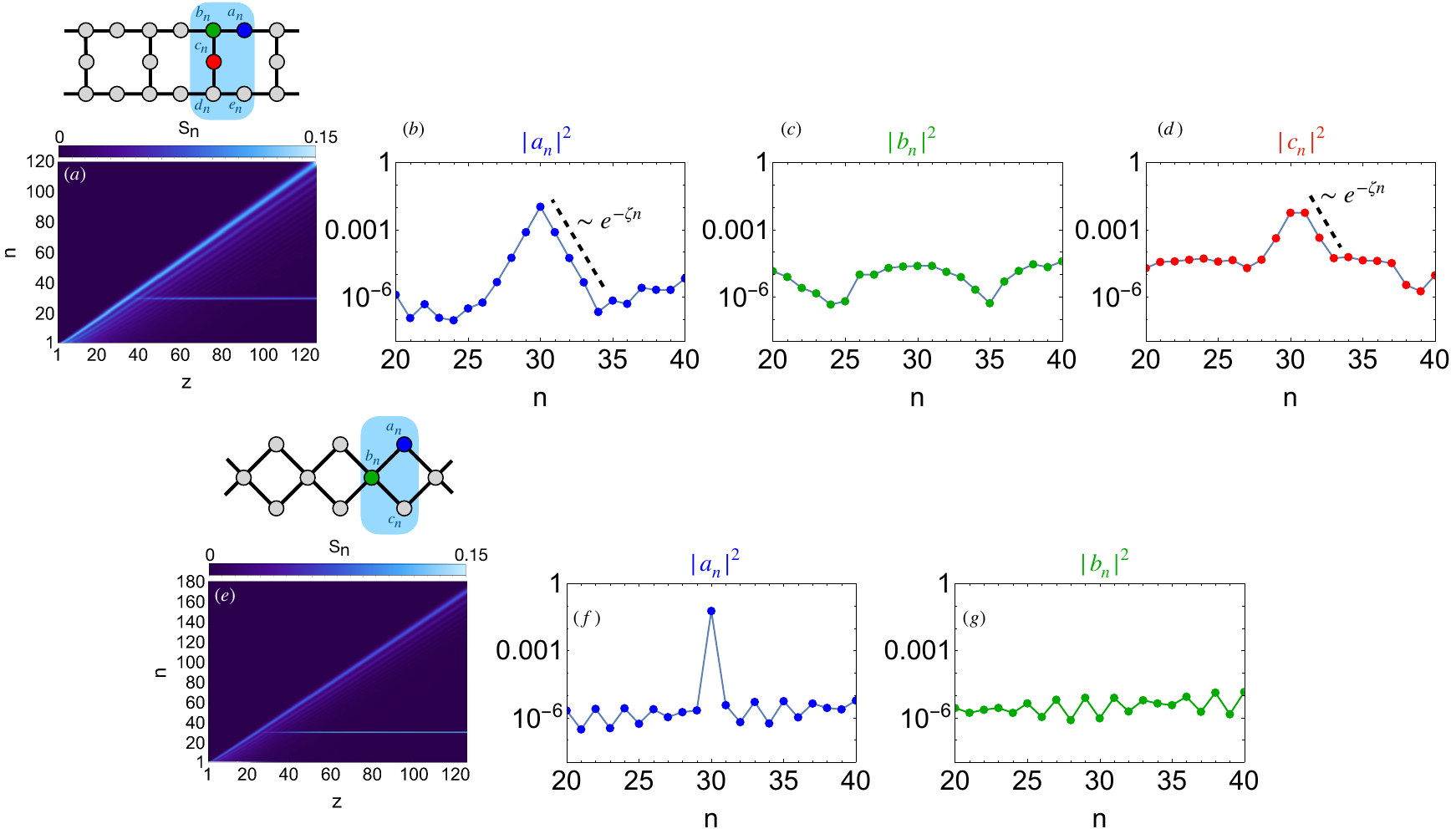}}
\caption{(a) Intensity profile evolution $S_n$ for an initial single-site excitation centred at the defect waveguide for the Lieb ladder.  
(b-d) Amplitude $|a_n|^2$ [panel (b)], $|b_n|^2$ [panel (c)] and $|c_n|^2$ [panel (d)] as function of the unit-cell number $n$ at final propagation distance.  
(e-g) Same as (a-c) for the diamond chain. 
} 
\label{fig:state_LogLin}
\end{figure}

Let us consider an excitation at unit-cell $n=30$ of the Lieb ladder shown in Fig.~\ref{fig:state_LogLin}(a) generated by an incoming plane wave generated via the PW method~\cite{SM_Real2024Radiation-basedLattices}. In Fig.~\ref{fig:state_LogLin}(b-d) we plot the amplitudes in sites $a_n$, $b_n$ and $c_n$ respectively in linear-logarithmic scale. 
The intensities $|a_n|^2$ [panel (b)] and $|c_n|^2$ [panels (d)] as well as $|e_n|^2$ (not shown here as  analogous to $|a_n|^2$) report hints of exponential decay before reaching the background amplitude $\sim 10^{-6}$. The amplitude  $|b_n|^2$ [panel (c)], as well as $|d_n|^2$ (not shown here as analogous to $|b_n|^2$), instead essentially sit at the background amplitude $\sim 10^{-6}$.
Differently to the Lieb ladder where the CLS form a non-orthogonal set, in the diamond chain Fig.~\ref{fig:state_LogLin}(e) where the CLS are orthogonal this exponential decay is absent. 
Indeed, as shown in Fig.~\ref{fig:state_LogLin}(f,g) the amplitudes $|a_n|^2$ [panel (f)] as well as $|c_n|^2$ (not shown here) immediately drops to the background amplitude $\sim 10^{-6}$. 

\subsection{Velocity dependence of waveguides}

The femtosecond writing technique~\cite{SM_szameit_discrete_2005} consists on the focusing of ultra-short pulses, in our case with a pulse width of $~\sim 210$ fs and a wavelength of $1030$ nm, inside a glass material (Borosilicate in our experiments). The laser gives enough energy to the glass molecules to produce a local reordering of them, with an effective increment of the density. This modifies the electric susceptibility at the focal region, implying a direct modification of the refractive index contrast in the order of $10^{-4}-10^{-3}$, over the refractive index of the Borosilicate $n_0=1.48$. There are two crucial parameters in this technique that control the waveguides properties: the writing power $P_w$ and the writing velocity $v_w$. The standard recipe in the fabrication protocol indicates that a larger energy dose produces a stronger change in the material and, therefore, larger $P_w$ and slower $v_w$ are normally the right directions. However, the guiding properties and optical functionality depends on these parameters, which is also strongly affected by the excitation wavelength. For example, by simply fabricating stronger waveguides, manifesting a high confinement, they become multimode easily~\cite{SM_SPdimer}. In addition, the present investigation demands us to have a good control over the fabrication parameters such to precisely determine the strength of the asymmetric potential. 

\begin{figure}[h!]
  \center{\includegraphics[width=10cm]{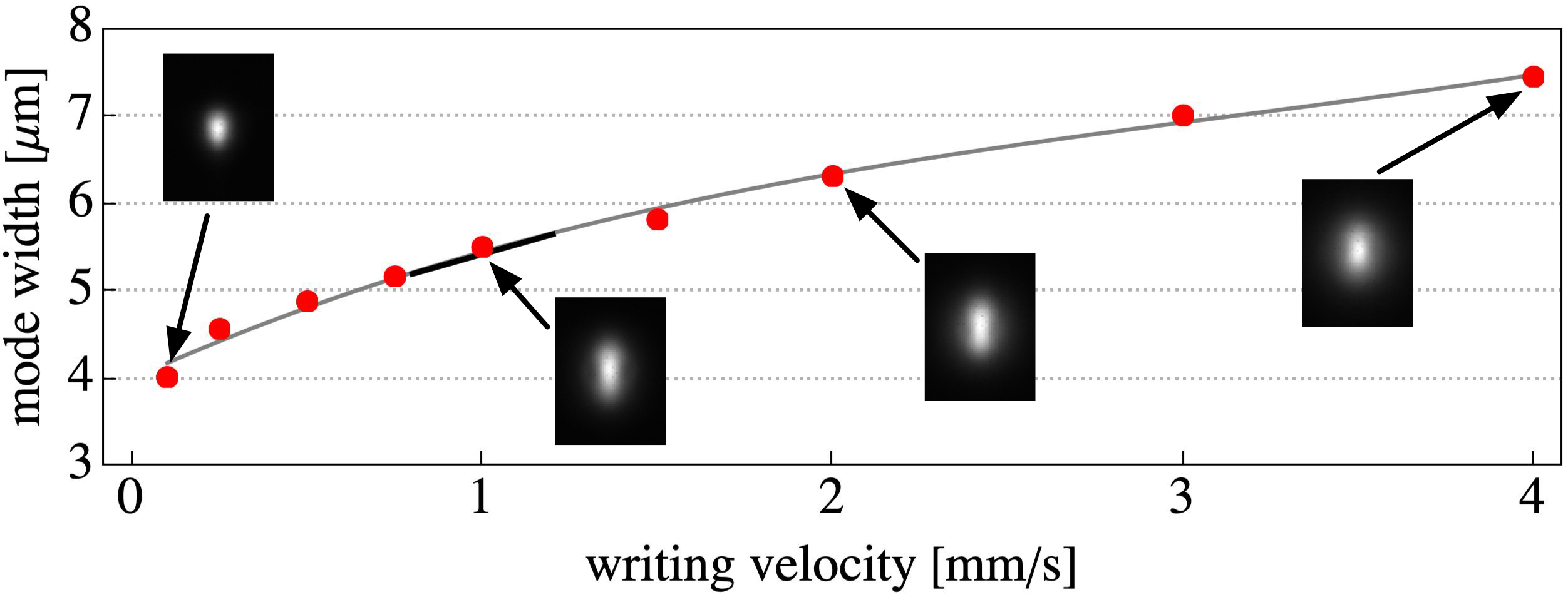}}
\caption{Horizontal width variation of guided modes versus writing velocity for $\lambda=750$ nm. Red disks show the experimentally extracted horizontal widths. The gray curve corresponds to a third order polynomial fit and the black straight line to a linear dependence in the region of interest. Insets show examples of guided modes intensity profiles at the indicated velocities. } 
\label{smwvelo}
\end{figure}

Considering all these constraints, we decided to fabricate single mode waveguides for an excitation wavelength in the range $\lambda\in\{700,790\}$ nm. For this we chose a nominal fabrication power $P_w=20.52$ mW and we explored the response of the guided modes with respect to the fabrication velocity $v_w$. Fig~\ref{smwvelo} shows the dependence of the guided mode width, extracted after fitting the experimental profiles with $\sinh$-like functions. We observe a monotonous reduction of the horizontal width while the waveguide is fabricated with a slower velocity. We found a good linear region around $v_w=1$ mm/s, where we observe a relatively large slope on a narrower region. This allowed us to introduce small changes in the asymmetric potential such to carefully calibrate the right experimental conditions for observing the trapping of FB localized states, as it is described in the Main text and below.

\subsection{Dimer experiments and coupling constant determination}

\begin{figure}[h!]
  \center{\includegraphics[width=14cm]{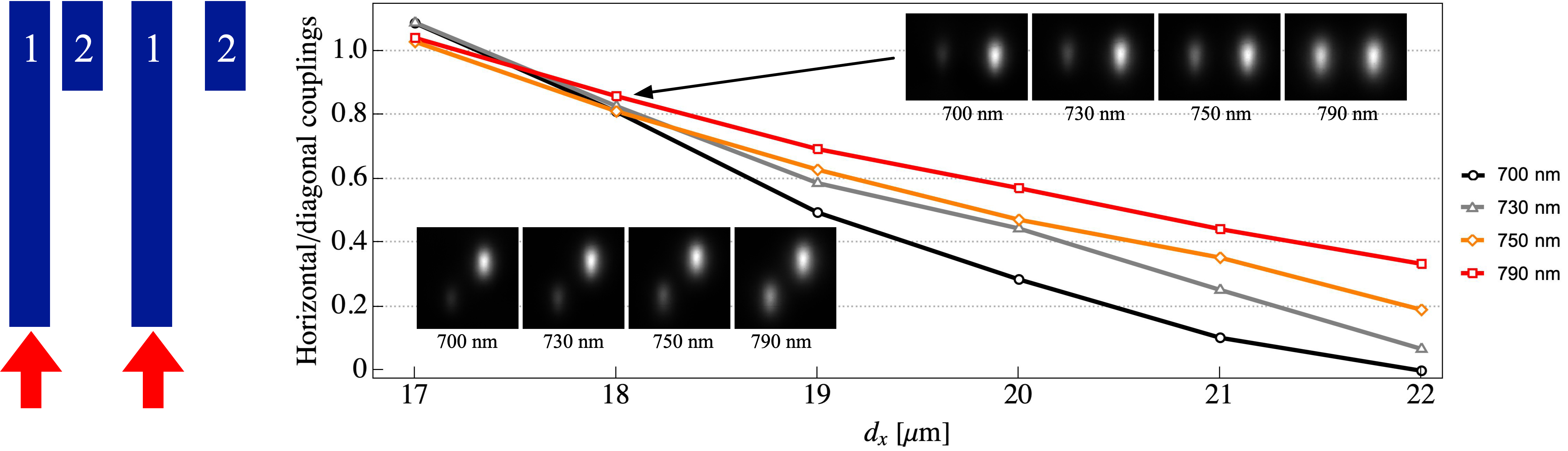}}
\caption{Left: Dimer experiment configuration. Right: Experimental determination of coupling constants. Symbols show the ratio in between the extracted horizontal coupling constants divided by the diagonal ones, for a fixed diagonal distance of $18.4\ \mu$m and different horizontal distances $d_x$. The upper insets show the intensity output profiles at $d_x=18\ \mu$m for the horizontal couplers. The bottom insets show the intensity output profiles of a diagonal coupler for $d=18.4\ \mu$m.}
\label{smwcouplings}
\end{figure}

Before fabricating a lattice system, we calibrate the coupling constants by fabricating a set of dimers, as sketched in Fig~\ref{smwcouplings}-Left. A long waveguide (1) is fabricated along the glass length. A short waveguide (2) is fabricated at a given separation distance $d_x$, which has a typically length of $l=5-7$ mm. As the coupling constants in our experiments are of the order of $V=1$ cm$^{-1}$, then the second waveguide is shorter than the coupling length $l_c=\pi/2C$. Therefore, in this way, we are able to measure the output intensities and estimate the coupling constants using the standard formula

\[V(d_x)=\frac{1}{l}\tan^{-1}\sqrt{\frac{P_2(l)}{P_1(l)}}\ ,\]
which comes from the dynamical solution of a photonic dimer after single-site excitation~\cite{SM_nore25}.

Fig.~\ref{smwcouplings}-Right shows the collected results after extracting the intensities profiles for each photonic coupler. We specifically compare the horizontal coupling versus the intersite distance $d_x$ divided by the diagonal coupling, obtained for a diagonal distance of $18.4\ \mu$m. We show the data obtained for 4 different excitations wavelengths. We observe a coupling ratio smaller than $1$ for distances larger than $17\ \mu$m, what is one of the conditions required for the satisfactory excitation of a plane wave with quasi-momentum $k_x=\pi/2$~\cite{SM_Real2024Radiation-basedLattices}.

\subsection{numerical PW excitation in 1D, diamond and 1D Lieb, including fourier}

Now, we study numerically the method of generating PW with a defined quasi-momentum $k_x=\pi/2$. The concept comes from a previous work~\cite{SM_Real2024Radiation-basedLattices}, in which an edge-coupled defect produces the appearance of a stationary impurity state having a large amplitude at the defect plus a long and constant tail with a well defined phase structure. This phase is fully determined by the detuning of the defect site with respect to the lattice. For a zero detuning, the impurity mode has a frequency/energy equal to zero and, therefore, it has a tail with a $\pi/2$ phase structure. So, when exciting the system by injecting light at the defect site, first of all a large peaked profile is generated, where the size of this peak compared to the long tail structure directly depends on the coupling of this site and the lattice: $V_{pw}$. Then, as this input excitation mostly excites the impurity state, the dynamics develops and increasing tail which looks as a gaussian profile moving away from the defect. This beam has a well defined phase structure and can be used as a PW generator for the studied lattice~\cite{SM_Real2024Radiation-basedLattices}.

\begin{figure}[h!]
  \center{\includegraphics[width=14cm]{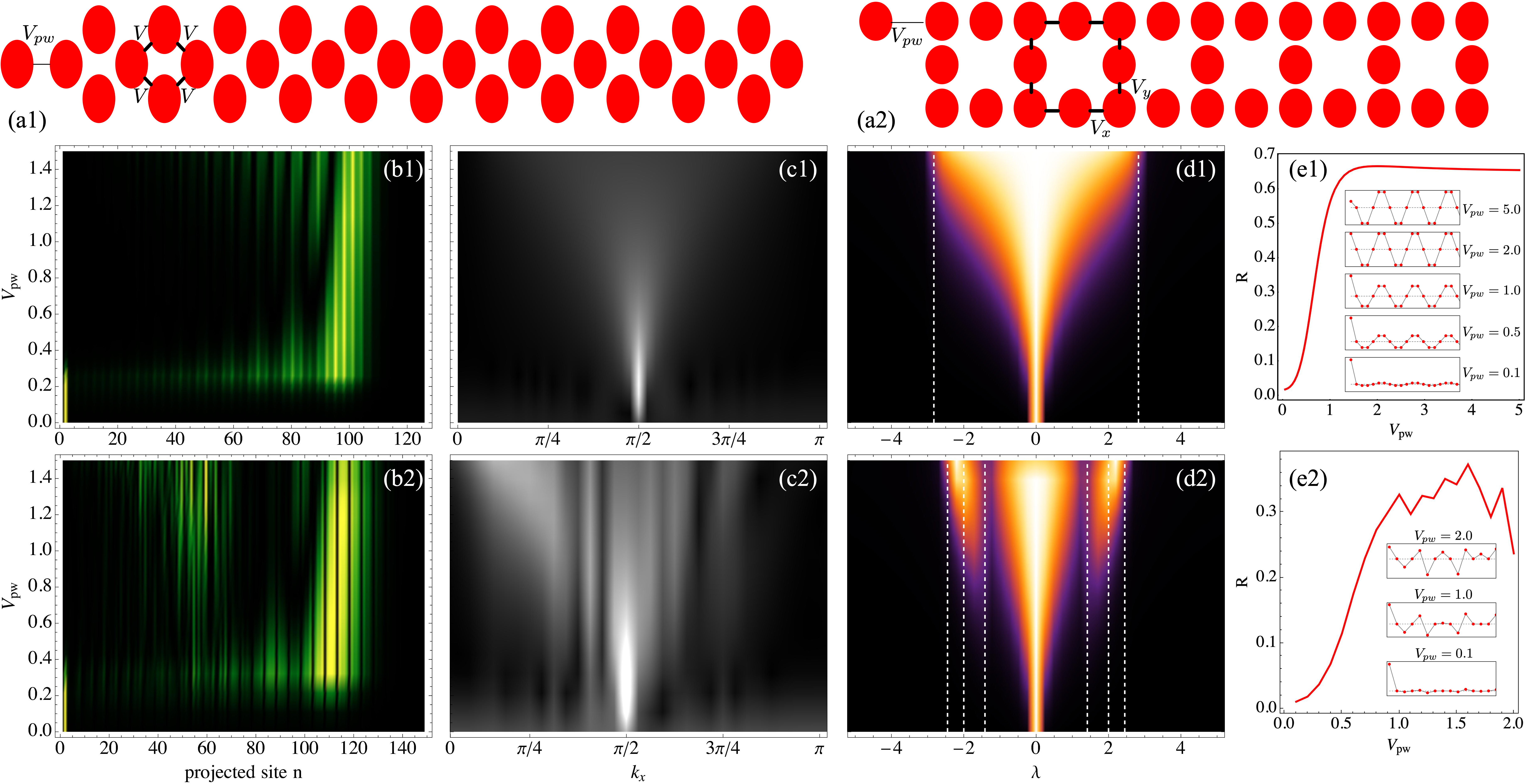}}
\caption{Numerical study of PW generators for Diamond (1) and 1D Lieb (2) lattices. (a) Geometries under investigation. (b) Projected output intensity profiles at $z=25$ versus $V_{pw}$. (c) Transversal Fourier transform $k_x$ at $z=25$ versus $V_{pw}$. (d) Longitudinal Fourier transform $k_z$ ($\lambda$) versus $V_{pw}$. Dashed lines in (d) show the bands edges. (e) Participation ratio $R$ versus $V_{pw}$ for stationary edge states at $\lambda=0$. Insets in (e) show examples of projected amplitude profiles at the indicated values of $V_{pw}$.} 
\label{smSimsPW}
\end{figure}

We explore this concept for the Diamond and 1D Lieb lattices, with the geometries sketched in Figs.~\ref{smSimsPW}(a1) and (a2), respectively. We start by exploring the PW generator method in Diamond lattices. In Fig.~\ref{smSimsPW}(b1) we show the projected output intensity profiles obtained at $z=25$, for a lattice with coupling constant $V=1$. We observe that the propagating profile to the right is simpler around $V_{pw}\approx 0.7$. That means that essentially a single gaussian-like profile is traveling through the lattice. For smaller values of $V_{pw}$, we observe a strongly localized profile having a large peak at the defect site with a rather small (negligible) tail. Then, we compute the transversal Fourier transform and extract the $k_x$'s values of the propagating profile [see Fig.~\ref{smSimsPW}(c1)]. We observe a quite narrow peak around $k_x=\pi/2$ for $V_{pw}\lesssim 0.8$. Of course, a narrower profile in $k_x$-space is an important goal in order to have a well-defined PW. However, we also observe that this occurs only for very small $V_{pw}$, what naturally affect the possible experiment as the effective part of the beam traveling through the lattice will be very small. In addition, it will take a larger propagation distance to evacuate the injected energy from the defect into the lattice due to the weak coupling. Both restrictions are indeed quite relevant for the experiment. We compute a longitudinal Fourier transform~\cite{SM_UtaSTT13} to extract the longitudinal propagation constants ($k_z$ or $\lambda$) which are excited during the dynamics, after a given input condition. Fig.~\ref{smSimsPW}(d1) shows the spectrum excited during the dynamics shown in Fig.~\ref{smSimsPW}(b1), where we observe a clear transition from a pure $\lambda=0$ state, corresponding to the defect mode described above, and an increasing excited spectrum which get wider for $V_{pw}\gtrsim 0.8$. For $V_{pw}\gtrsim 1.0$ we expect to excite the whole band structure, excepting the flat bands. Finally, we compute the localized stationary states at a frequency $\lambda=0$ decaying from the defect site at the left edge [see Fig.~\ref{smSimsPW}(e1)]. The inset profiles show that these states decay into the lattice for $V_{pw}<2.0$, with a clear peaked solution. We observe that these states increase their effective size, determined by the participation ratio $R\equiv (\sum |u_{\vec{n}}|^2)^2/(\sum |u_{\vec{n}}|^4)$, while $V_{pw}$ increases, indicating a growing tail. Therefore, dynamically speaking, when the amplitude at the tail is comparable to the amplitude at the defect site, a single-site excitation there is less effective as it also excites other states having an amplitude different to zero at that site.

The 1D Lieb lattice sketched in Fig.~\ref{smSimsPW}(a2) shows an extra site as PW generator at the top of the lattice, but of course this could exist also at the bottom. This lattice is much more complicated in terms of PW generation due to the larger number of possible couplings in the dynamics. However, Fig.~\ref{smSimsPW}(b2) shows the possibility of generating a propagating beam through the lattice for $V_{pw}\lesssim 0.8$, with a larger and faster main wavepacket. The transversal Fourier spectrum shown in Fig.~\ref{smSimsPW}(c2) is quite noisy in this case, but there is a larger peal at $k_x \approx\pi/2$. The longitudinal frequency spectrum shown in Fig.~\ref{smSimsPW}(d2) describe a rather clean excitation of a $\lambda=0$ states for $V_{pw}\lesssim 0.8$, although we can also observe weaker excited modes closer to the band edges. The stationary defect states at $\lambda=0$ described in Fig.~\ref{smSimsPW}(e2) show a similar behavior than the ones for the Diamond lattice, with an increasing effective size for larger $V_{pw}$.

\subsection{Experimental PW optimization in diamond and Lieb homogeneus lattices}

\begin{figure}[h!]
  \center{\includegraphics[width=14cm]{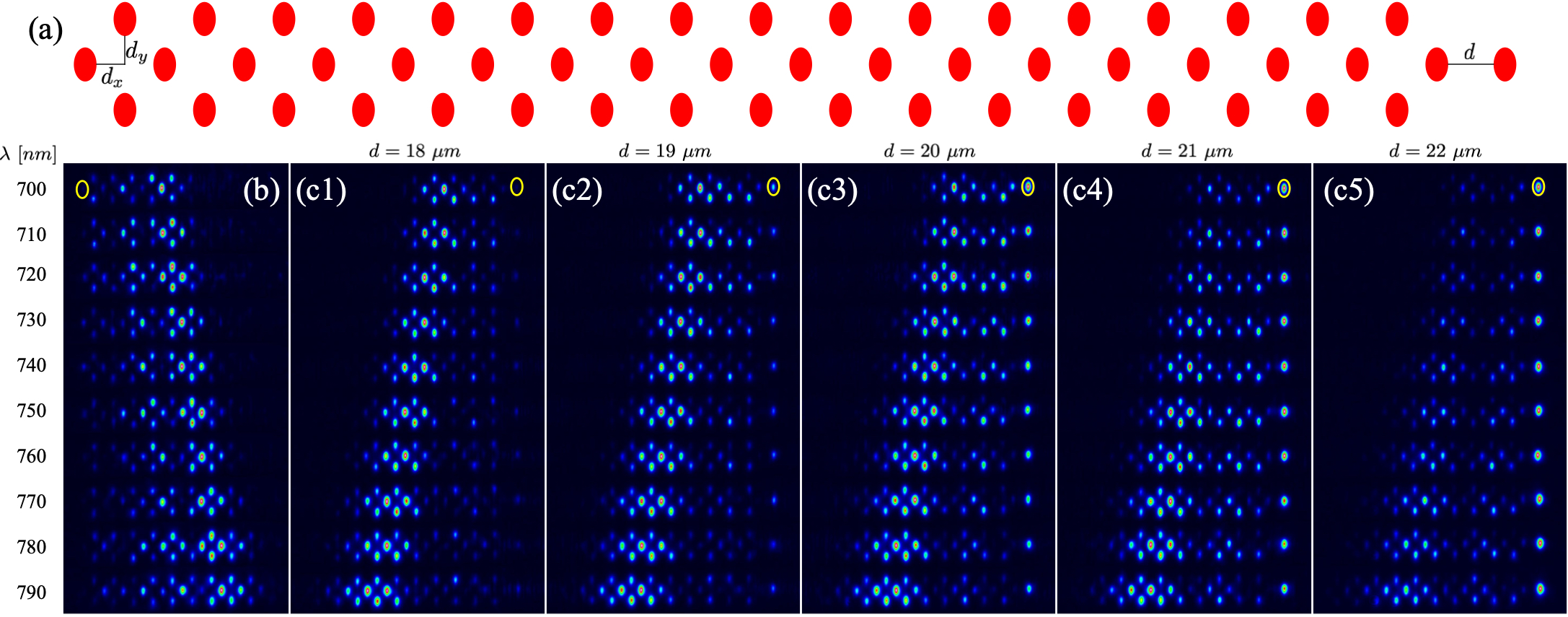}}
\caption{(a) Diamond geometry. Comparison between (b) discrete diffraction and (c) PW generators for a Diamond lattice with $d=13\ \mu$m, and for different excitation wavelengths as indicated at the left, and for the PW generator distance $d_x$. The yellow ellipse indicates the input excitation position.} 
\label{smPWdia}
\end{figure}

\begin{figure}[h!]
  \center{\includegraphics[width=14cm]{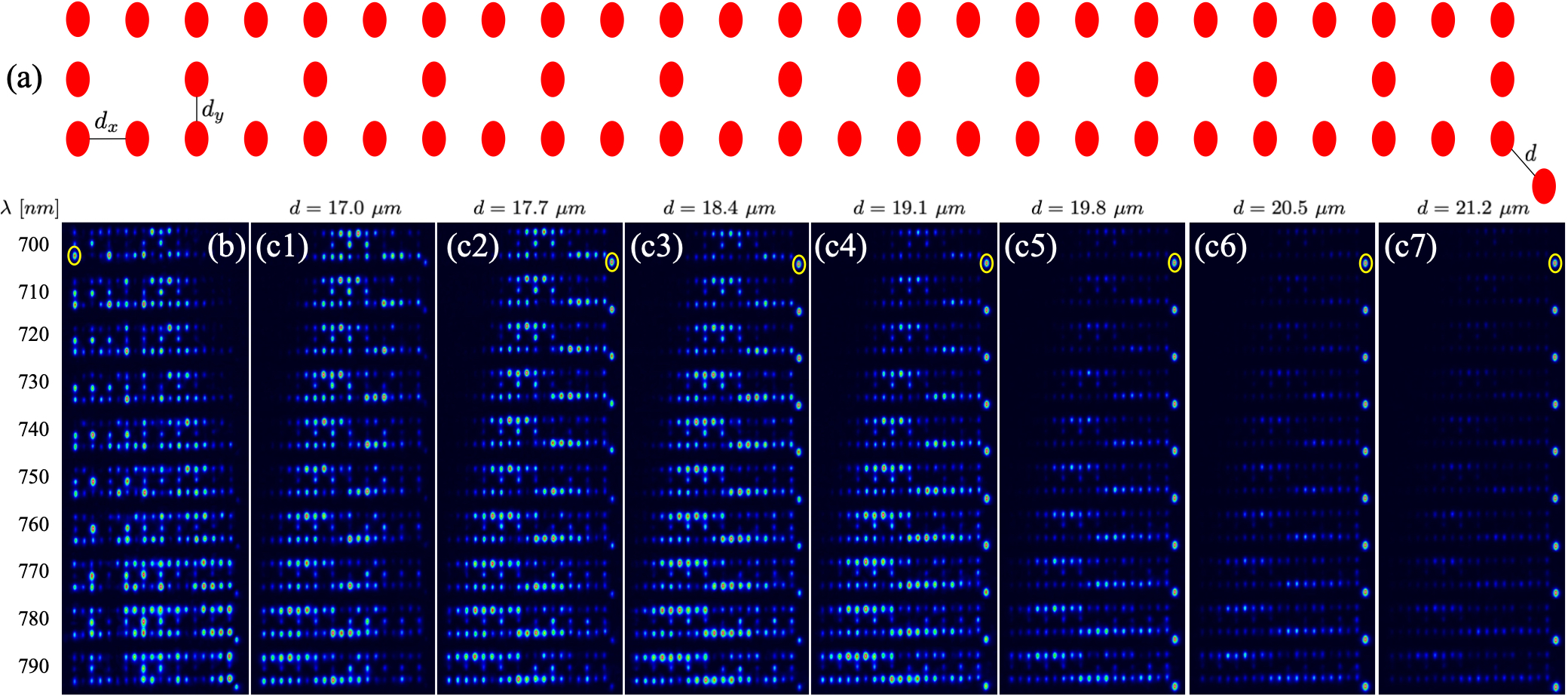}}
\caption{(a) 1D Lieb geometry. Comparison between (b) discrete diffraction and (c) PW generators for a 1D Lieb lattice with $d_x=15\ \mu$m and $d_y=19\ \mu$m, for different PW diagonal distances $d$, and for different excitation wavelengths (indicated to the left). The yellow ellipse indicates the input excitation position.} 
\label{smPWlieb}
\end{figure}

We explore the PW generation experimentally with the geometries shown in Figs.~\ref{smPWdia}(a) and \ref{smPWlieb}(a). We fabricate homogeneous Diamond and 1D Lieb lattices with different defect sites as PW generators, and study the effective dynamics, of a fixed lattice, by sweeping on the excitation wavelength~\cite{SM_diamondAPL23,SM_graphene24,SM_2Drad25}. We first study the discrete diffraction pattern produced by exciting a single site at the left edge of both lattices [see Figs.~\ref{smPWdia}(b) and ~\ref{smPWlieb}(b)]. This input condition excites several extended states of the lattice and, therefore, demonstrate the standard dispersion properties of both lattices. We observe a simple diffractive pattern (quite similar to a 1D lattice~\cite{SM_repLederer}) for Diamond, while a more noisy pattern for Lieb.

Then, we explore the excitation of the PW generators by increasing the distance of the defect site with respect to the lattices, which produces a reduction of $V_{pw}$. For Diamond lattices [Fig.~\ref{smPWdia}(c)], we observe that there is a well defined profile for $d=18,19\ \mu$m, that is moving away from the defect site at the right edge. Then, we observe that the excited profile has a large peak at the edge site, with a more noisy diffractive pattern. This coincides with the numerical results observed in Fig.~\ref{smSimsPW}(b1) for smaller $V_{pw}$.

For 1D Lieb lattices [Fig.~\ref{smPWlieb}(c)], we observe a more complex pattern with clear diffractive fronts traveling along the upper and bottom rows, but somehow disconnected. This is indeed a problem in the method, but it can not be solved in a simple way, for example by adding a second defect site, as a stronger localized state is generated in such configuration. Nevertheless, we observe a diffractive pattern that propagates through the lattice. From the simulations shown in Figs.~\ref{smSimsPW}(b2)-(d2), we know that this pattern is a mixture of linear modes around $\lambda=0$, which the necessary requirement for the excitation of the FB at the asymmetric potential. Similar to the Diamond case, we observe that the intensity of the diffractive pattern is reduced while decreasing $V_{pw}$. At the same time, the amplitude at the defect site increases, as expected from the stationary solutions shown in Fig.~\ref{smSimsPW}(e2).

\subsection{Experimental PW optimization for Diamond lattices}

With all the previous information, we started the optimization of the experimental conditions to achieve the excitation of FB states through PW. A first step was to find a right geometry for the Diamond lattice. In principle, any fabricated lattice will be well described by tight-binding like models if the intersite distances are large enough to avoid next-nearest neighbor couplings. This, of course, also depends on the excitation wavelength (for larger colors, the guides profiles are wider and the superposition stronger) and the polarization (horizontally polarized guided modes are wider than the ones observed for vertical polarization). Therefore, experimentally speaking, it is mandatory the adjustment of the lattice geometry such to observe a clear discrete dynamics and as much propagation as possible, for a fixed glass length. In the specific Diamond case, we simply took some optimal geometry found in previous experiments, with equal horizontal and vertical distances $d_x=d_y=13\ \mu$m [see Fig.~\ref{smPWdia}(a)], giving a diagonal nearest neighbor distance of $18.4\ \mu$m. Typically, an inter-site distance larger than $15\ \mu$m shows a very good agreement of the coupled mode theory~\cite{SM_nore25} and our experiments.

\begin{figure}[h!]
  \center{\includegraphics[width=12cm]{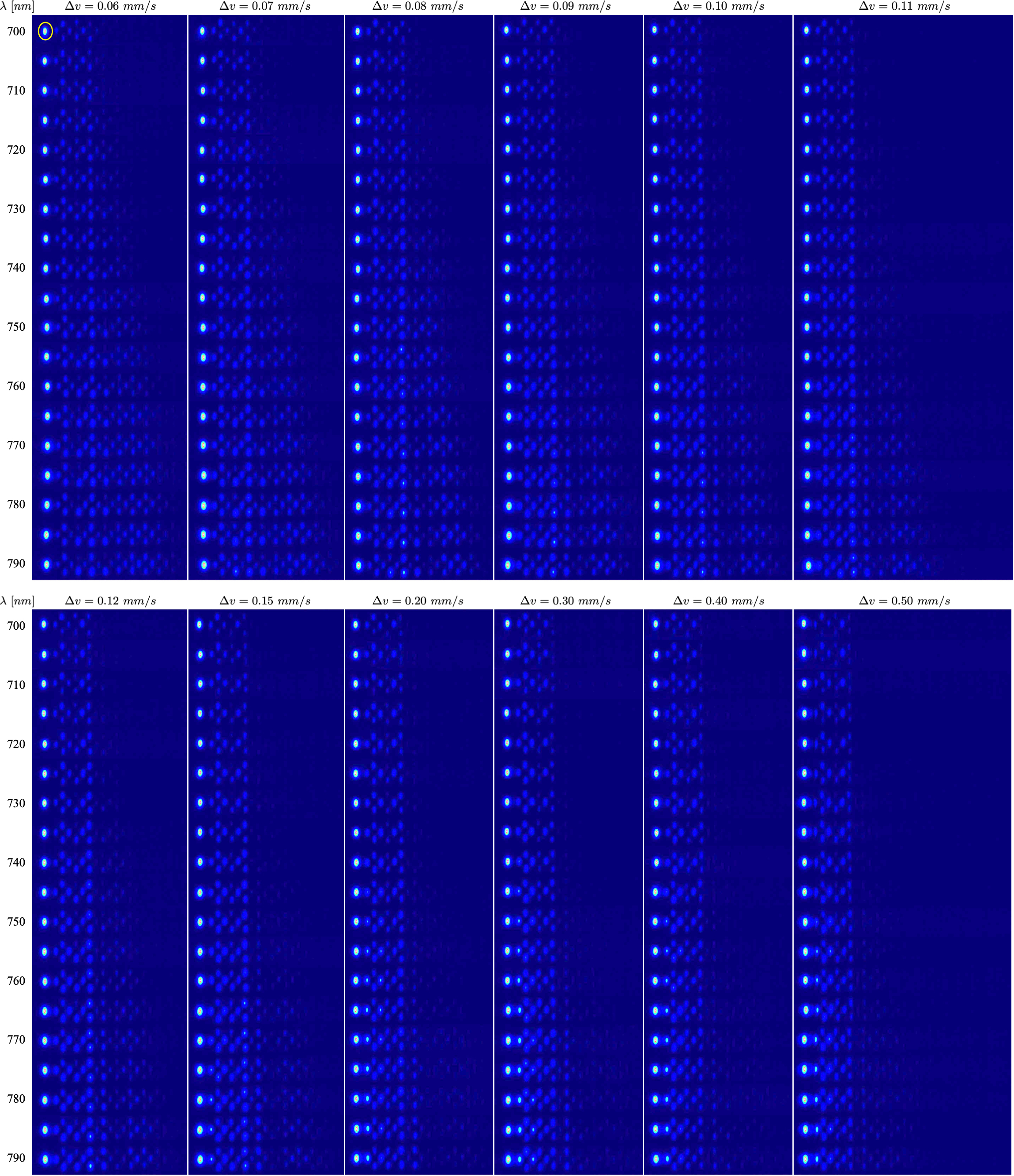}}
\caption{Tuning of the potential strength. PW excitation of the FB state at the third plaquette of a diamond photonic lattice, for different excitation wavelengths as indicated at the left. The lattice geometry is given by $d_x=d_y=13\ \mu$m, and a PW generator located at a horizontal distance $d=22.0\ \mu$m. All the waveguides are fabricated with a writing velocity $v_w=1.0$ mm/s. The asymmetric potential is inserted by modifying the upper and bottom waveguides writing velocities of a given plaquette by an amount $\Delta v$, as indicated at the top of each panel. The asymmetric potential has been cut at $z=43$ mm, for a total propagation of $48$ mm. The yellow ellipse indicates the input excitation position.} 
\label{smDiaStrength}
\end{figure}

The first calibration we made in this problem was with respect to the strength of the asymmetric potential. As it is shown in Fig.3(a) of the main text, we insert this potential by applying a velocity gradient in the potential region. The lattice is fabricated with a velocity $v_{w}=1.0$ mm/s and power $P_w=20.5$ mW, and the asymmetric potential waveguides are fabricated with velocities $v_w+\Delta v$ and $v_w-\Delta v$ for the upper and bottom waveguides, respectively. Therefore, $\Delta v$ determines the strength of the potential, in a region where the changes of the guiding properties can be assumed as linear [see Fig.~\ref{smwvelo}]. Fig.~\ref{smDiaStrength} shows the effect of varying $\Delta v$ at the third plaquette. The intensity profiles are obtaining by sweeping on the excitation wavelength, such to observe an effective dynamics. Additionally, we have saturated the images quite a lot, such to observe the background waves excited for the PW generator, which in this specific case was set as $d=22\ \mu$m. For weaker potentials, we observe that some part of the energy is able to pass through the third plaquette and also some part remains at the left edge. For example, for $\Delta v=0.08,0.09$ mm/s we observe a tendency to trap the energy at the third plaquette, with approximately one half of the energy been reflected and one half been transmitted. Then, for larger $\Delta v$ we observe a notorious reduction of the transmitted light with an increasing reflected pattern which interacts with the localized state formed at the third plaquette. This of course affects the excitation of the FB state at the third plaquette, as the reflected waves keep there unstabilizing the compact profile. For very large $\Delta v$ we observe almost no transmitted light and, also, no FB state formed. With these resutls, we found an optimal regime for the region around $\Delta v\approx 0.08$ mm/s.

\begin{figure}[h!]
  \center{\includegraphics[width=14cm]{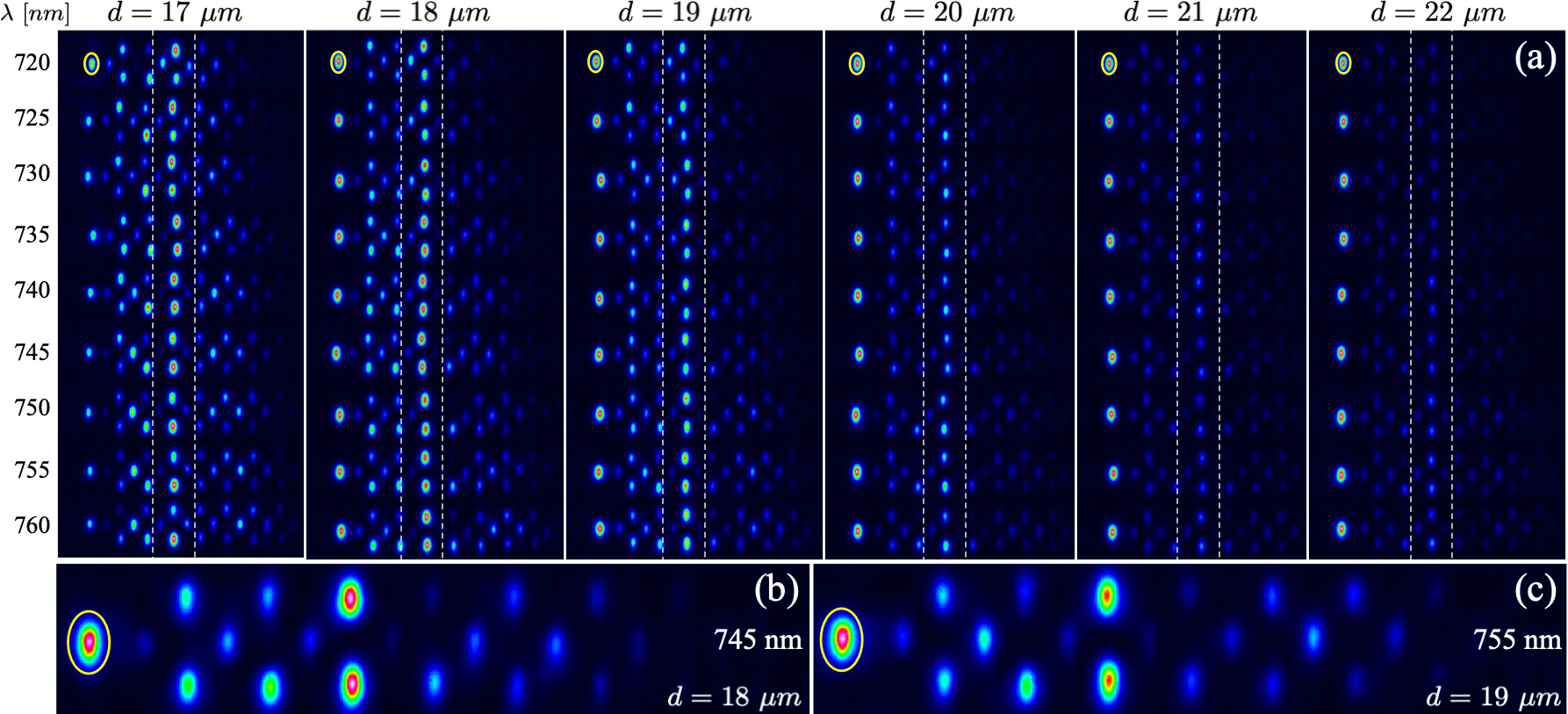}}
\caption{Optimization of the PW generator for Diamond lattices. (a) PW excitation of the FB state at the third plaquette of a diamond photonic lattice, for different excitation wavelengths as indicated at the left. The lattice geometry is given by $d_x=d_y=13\ \mu$m, and the PW generator is located at a horizontal distance indicated at the top of each panel. The writing velocities of all the waveguides is $v_w=1.0$ mm/s, excepting the ones of the asymmetric potential given by $\Delta v=0.08$ mm/s, which has been cut at $z=43$ mm, for a total propagation of $48$ mm. (b) and (c) Examples of intensity output profiles at the indicated parameters. The yellow ellipse indicates the excitation position.} 
\label{smDiaPWexp}
\end{figure}

Then, a second calibration is related to the optimal value for the PW distance $d$ in a Diamond lattice. In the previous figure~\ref{smDiaStrength}, the expected phenomenology was observed, but with a strong impurity profile having a very large amplitude at the defect site. Therefore, we study now the effect of varying the PW generator distance $d$ as it is described in Fig.~\ref{smDiaPWexp}(a), all for $\Delta v=0.08$ mm/s. We observe an interesting emergence of the FB localized state, at the third plaquette, for all the cases. However, we observe a more symmetric and homogeneous excitation for $d=18\ \mu$m, over a large range of wavelengths, which also coincides with a very good PW generation [see Fig.~\ref{smPWdia}(c1)]. For $d=17\ \mu$m we observe a more noisy pattern at the left of the third plaquette, while for $d\geq19\ \mu$m the FB localized state is clearly excited but with a very small intensity profile, due to the increment of the amplitude at the defect site, what is consistent with the numerical results described in Fig.~\ref{smSimsPW}. Figs.~\ref{smDiaPWexp}(b) and (c) show two examples for an almost perfect excitation of a FB at the third plaquette. The out of phase structure is evident by observing that in between the two main peaks we see a very dark (black indeed) color. This means a discontinuity in the phase of the upper and bottom amplitudes, passing through a zero amplitude value, as it is expected for a FB state excitation. All the data presented in Fig.~\ref{smDiaPWexp} has been obtained after optimizing the length of the asymmetric potential, as described in the main text.

\subsection{Experimental PW optimization for 1D Lieb lattices}

\begin{figure}[h!]
  \center{\includegraphics[width=16cm]{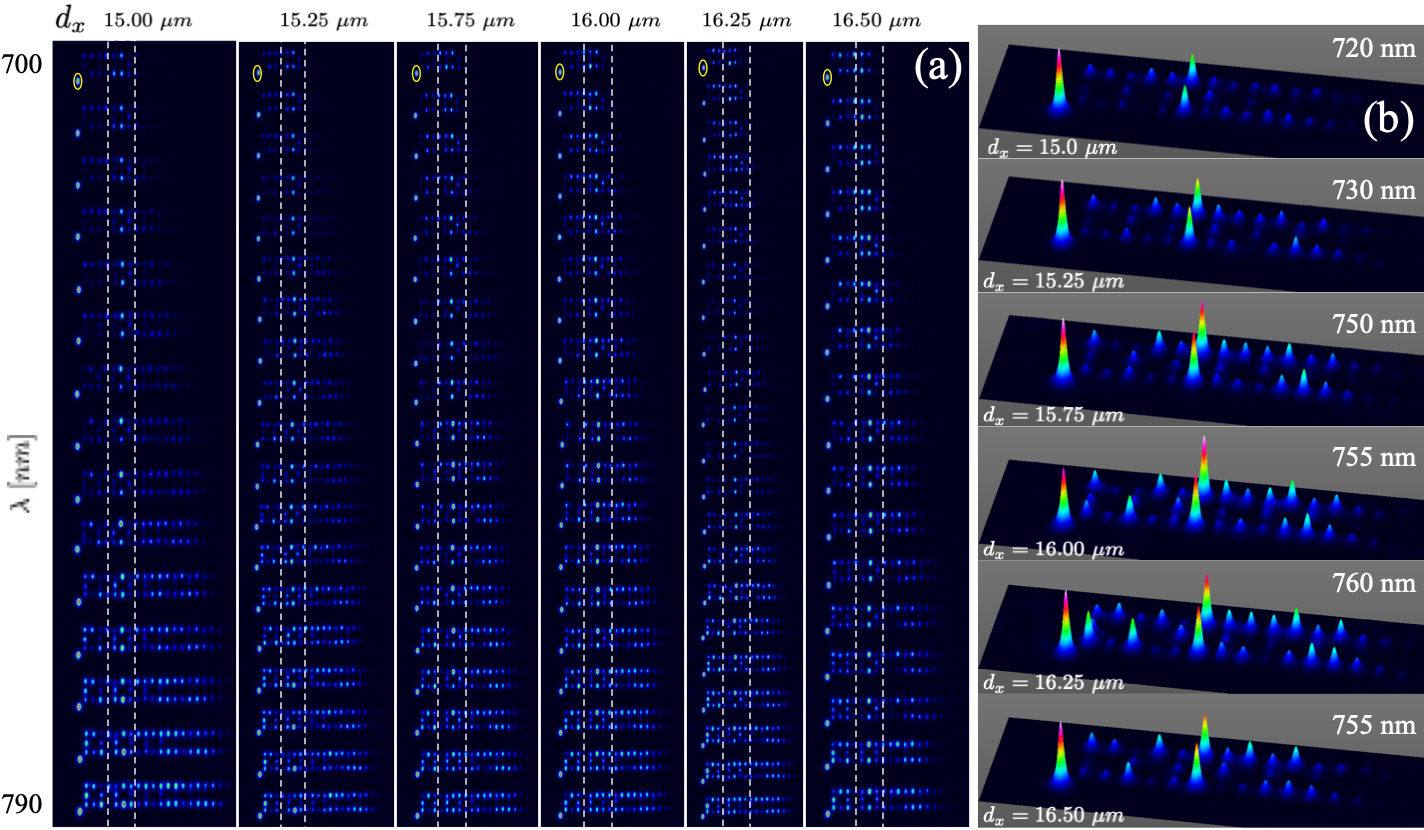}}
\caption{Horizontal lattice dilatation. (a) PW excitation of the FB state at the third plaquette of a 1D Lieb photonic lattice, for different excitation wavelengths as indicated at the left. The lattice geometry is defined by a horizontal distance $d_x$ indicated at the top of the figure, a vertical distance $d_y=17\ \mu$m, and a PW generator located at a diagonal distance $d=18.4\ \mu$m. (b) 3D profiles at the indicated wavelengths and $d_x$. The asymmetric potential has been cut at $z=43$ mm, for a total propagation of $48$ mm. The yellow ellipse indicates the input excitation position.} 
\label{smLiebdx}
\end{figure}

Now, we optimize the experimental conditions for observing a trapped state at the third plaquette of a 1D Lieb chain. This geometry is much more complicated, as a good balance of coupling constants is required such to observe good transport conditions but also a more symmetric FB state, which for a Lieb geometry has a four-site pattern~\cite{SM_Vicencio2015a,SM_1DLieb}. However, in general, the PW method in 1D Lieb shows the excitation of quite asymmetric profiles, with the larger peaks at the top and bottom sites (and weaker at the left and right ones). Fig.~\ref{smLiebdx}(a) shows a first calibration of the horizontal Lieb geometry after varying the inter-site distance $d_x$, while keeping the vertical distance $d_y=17\ \mu$m fixed [see the geometry in Fig.~\ref{smPWlieb}]. We observe that by increasing $d_x$ the FB state excited remains quite asymmetric and, also, the horizontal propagated waves reduce their transversal extension due to the reduction of the horizontal coupling constant. Fig.~\ref{smLiebdx}(b) shows different examples of 3D intensity output profiles at different $d_x$ and excitation wavelengths. In all these examples, the excited FB state is quite asymmetric, but, also, the effect is quite clear as we are indeed able to observe a quite clear localized state at the potential region. Therefore, independently of the symmetry of this state and the complexity of the PW generator for 1D Lieb lattices, we are able to effectively excite energy/information at a wished and well defined position.

\begin{figure}[h!]
\center{\includegraphics[width=12cm]{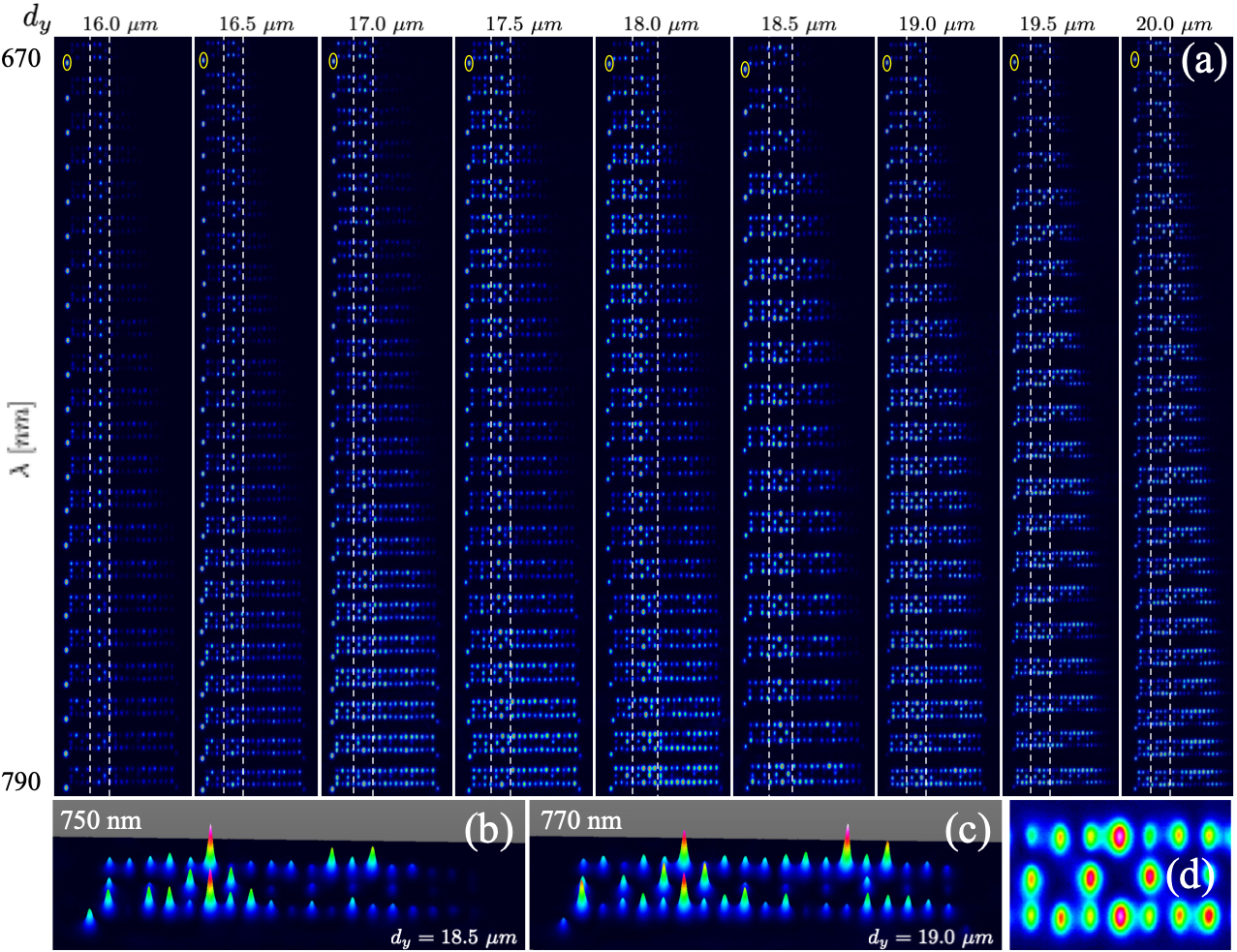}}
\caption{Vertical lattice dilatation. (a) PW excitation of the FB state at the third plaquette of a 1D Lieb photonic lattice, for different excitation wavelengths as indicated at the left. The lattice geometry is defined by a horizontal distance $d_x=15\ \mu$m, a vertical distance $d_y$ indicated at the top of the figure, and a PW generator located at a diagonal distance $d=18.4\ \mu$m. (b) 3D profiles at the indicated wavelengths and $d_y$. (d) Log scale intensity profile for (c). The asymmetric potential has been cut at $z=43$ mm, for a total propagation of $48$ mm. The yellow ellipse indicates the input excitation position.}
\label{smLiebdy}
\end{figure}

A second calibration consists on the variation of the vertical Lieb geometry $d_y$, while keeping the horizontal distance $d_x=15\ \mu$m fixed [see the geometry in Fig.~\ref{smPWlieb}]. In Fig.~\ref{smLiebdy}(a) we observe how a vertical elongation produces an interesting result with a FB state which is less asymmetric, although the upper and bottom peaks persist to be stronger than the left and right ones. We observe a very good and more symmetric four sites profile at $d_y\approx 19\ \mu$m, as shown in Figs.~\ref{smLiebdy}(b)--(d). It is very interesting to see that, although the PW generator method doesn't work perfectly for 1D Lieb lattice [see Fig.~\ref{smSimsPW}], we nevertheless excite states around a frequency $\lambda=0$, which are in a good resonant condition with the FB state we want to excite at the third plaquette. Also, we expect the excitation of the Lieb FB modes to be much more noisy, as neighbor FB modes share the left and right amplitude. So, this case is quite different to the Diamond lattices in which neighbor FB states do not share any amplitude.

\begin{figure}[h!]
\center{\includegraphics[width=12cm]{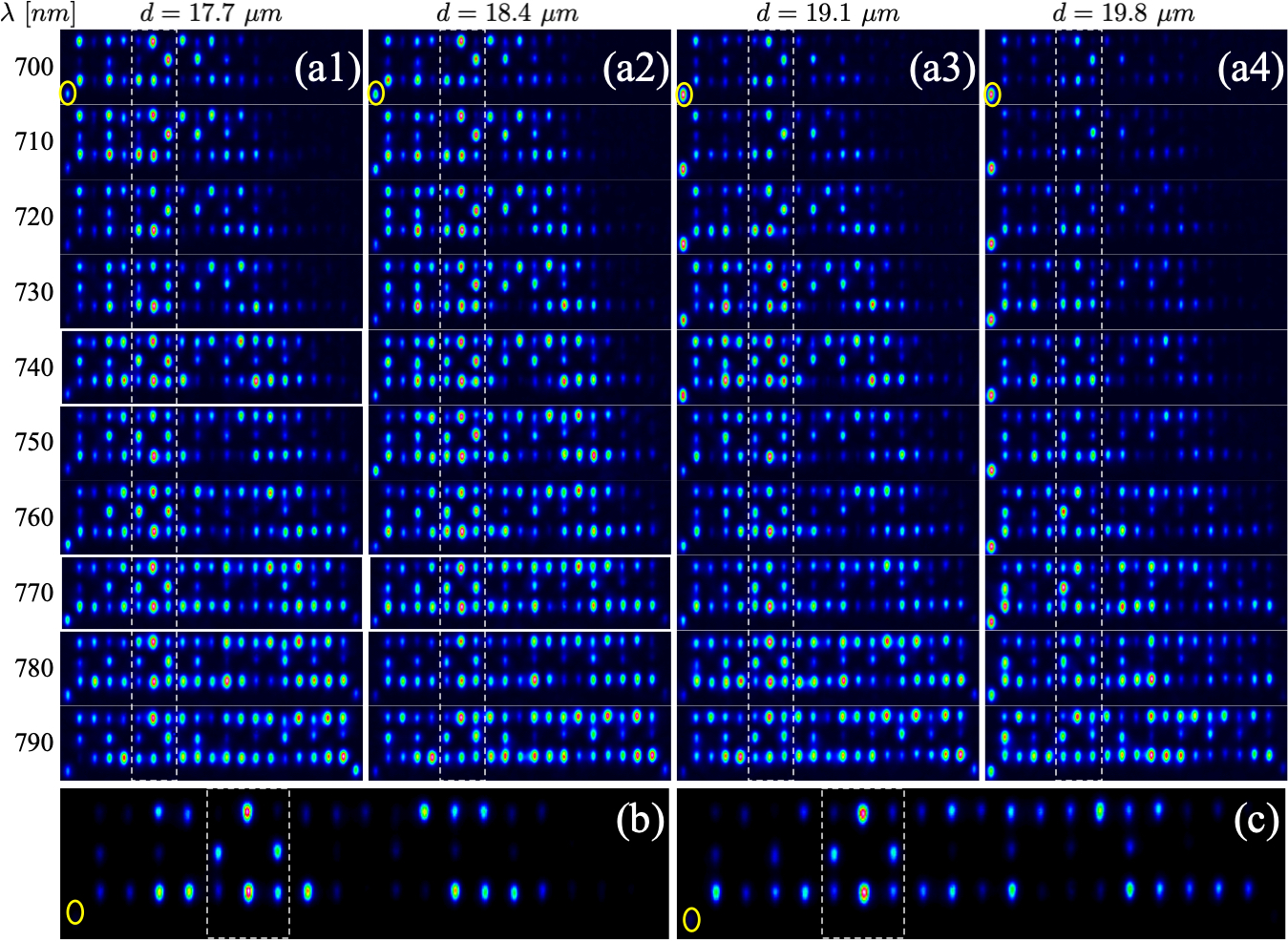}}
\caption{Final calibration of the PW generator distance $d$. (a) PW excitation of the FB state at the third plaquette of a 1D Lieb photonic lattice, for different excitation wavelengths as indicated at the left. The lattice geometry is defined by a horizontal distance $d_x=15\ \mu$m and a vertical distance $d_y=19\ \mu$m. (b) and (c) Square scaled intensity profiles for $d=17.7\ \mu$m ($\lambda=740$ nm) and $d=18.4\ \mu$m ($\lambda=770$ nm), respectively. The asymmetric potential has been cut at $z=43$ mm, for a total propagation of $48$ mm. The yellow ellipse indicates the input excitation position.}
\label{smLiebPWexp}
\end{figure}

After optimizing the geometry of the 1D Lieb lattice to $d_x=15\ \mu$m and $d_y=19\ \mu$m, we proceed to re-calibrate the PW generator distance $d$. Fig.~\ref{smLiebPWexp}(a) shows a short compilation of results for this final calibration, where we observe that a smaller distance and larger $V_{pw}$ produces a very clear FB state at the third plaquette, with an indeed quite symmetric profile. For increasing distances $d$ we observe a less intense excitation of the FB state, as the amplitude at the defect site increases and become predominant. Therefore, similar to Diamond lattices, we observe that the experimental excitation of FB states using the PW generator technique is more efficient for larger couplings $V_{pw}$, although a more precise beam in $k_x$-space should be obtained for smaller values of this coupling. Figs.~\ref{smLiebPWexp}(b) and (c) show two examples of intensity output profiles in a squared scale such to evidence the larger and more relevant amplitudes in the lattice. There, we can observe, quite clearly in both cases, the excitation of a compact localized FB state at the third plaquette of a 1D Lieb lattice.

The excitation of FB states using the asymmetric potential configuration and the PW generator setup will always demand a compromise in between quality and purity of the resonant excitation (at $k_x=0$ and $\lambda=0$), and the ability of the propagating beams to travel across the lattice. This last should be done as large as possible in the lattice such to efficiently excite the localized state, but also to propagate efficiently through the system. Of course, this is not a trivial balance as the glass length is fixed. We could be tempted to decrease the lattice dimensions to observe a larger transport, but FB states tend to be not robust to the presence of next nearest-neighbour interactions~\cite{SM_FBluis,SM_VicencioPoblete2021PhotonicDynamics}. Therefore, all the observations described in the present work required indeed a fine tuning process~\cite{SM_Danieli2024FlatApplications} to correctly generate efficient PWs, to calibrate the strength of the asymmetric potential, to find the optimal excitation wavelength and polarization, and the potential cut along the propagation direction (see discussion in the Main text).

\end{document}